\documentclass[prd,preprint,tightenlines,floatfix,showpacs,preprintnumbers,nofootinbib,eqsecnum,superscriptaddress]{revtex4}
\usepackage[cp1250]{inputenc} 
\usepackage[T1]{fontenc} 
\usepackage{amsfonts} 
\usepackage{amsopn}   
\usepackage{amssymb}  
\usepackage{amsthm}   
\usepackage{graphicx}
\usepackage{mathtools}
\usepackage{cancel}
\usepackage{enumerate}
\newcommand{\bq}{\mbox{\boldmath $q$}}
\newcommand{\br}{\mbox{\boldmath $r$}}

\newcommand{\bDelta}{\mbox{\boldmath $\Delta$}}

\newcommand{\bb}{\mbox{\boldmath $b$}}

\newcommand{\ket}[1]{| {#1} \rangle}
\newcommand{\bra}[1]{\langle {#1} |}

\newcommand{\half}{{1\over 2}}
\def\Pom{{\bf I\!P}}
\def\lsim{\mathrel{\rlap{\lower4pt\hbox{\hskip1pt$\sim$}}
		\raise1pt\hbox{$<$}}}         
\def\gsim{\mathrel{\rlap{\lower4pt\hbox{\hskip1pt$\sim$}}
		\raise1pt\hbox{$>$}}}         

\begin{document}

\vfill
\title{Coherent photoproduction of $J/\psi$ in nucleus-nucleus collisions in the color dipole approach}

\author{Agnieszka {\L}uszczak}
\email{agnieszka.luszczak@desy.de} 
\affiliation{
	T.~Kosciuszko Cracow University of Technology, PL-30-067 
	Cracow, Poland}

\author{Wolfgang Sch\"afer}
\email{Wolfgang.Schafer@ifj.edu.pl} \affiliation{Institute of Nuclear Physics Polish Academy of Sciences, 
	ul. Radzikowskiego 152, PL-31-342 Cracow, Poland}

\date{\today}

\begin{abstract}
We investigate the exclusive photoproduction of $J/\psi$-mesons in ultraperipheral
heavy ion collisions in the color dipole approach. We first test a number of
dipole cross sections fitted to inclusive $F_2$-data against the total cross section
of exclusive $J/\psi$-production on the free nucleon.
We then use the color-dipole formulation of Glauber-Gribov theory to calculate
the diffractive amplitude on the nuclear target. The real part of the free 
nucleon amplitude is taken into account consistent with the rules of Glauber 
theory. We compare our results to recent published and preliminary data 
on exclusive $J/\psi$ corrections in ultraperipheral lead-lead collisions
at $\sqrt{s_{NN}}=2.76 \, \rm{TeV}$ and $\sqrt{s_{NN}} = 5.02 \, \rm{TeV}$.
Especially at high $\gamma A$ energies there is room for
additional shadowing corrections, corresponding to triple-Pomeron terms
or shadowing from large mass diffraction. 
\end{abstract}

\pacs{}


\maketitle

\section{Introduction}

Following the early theoretical work \cite{Klein:1999qj} and the recent measurements
\cite{Abelev:2012ba,Abbas:2013oua,Khachatryan:2016qhq,Kryshen:2017jfz,LHCb:2018ofh}
(see also the review \cite{Contreras:2015dqa}) 
there has been recently much interest in the coherent exclusive 
production of vector mesons in ultraperipheral heavy-ion 
collisions at the LHC. The production takes place via the diffractive 
photoproduction process, where one of the ions serves as a source of quasireal
photons. The second ion plays the role of the hadronic target on which 
the diffractive photoproduction proceeds. 

The production of vector mesons composed of heavy quarks, such as the $J/\psi$
is of special interest and the exclusive production of $J/\psi$ 
in ultraperipheral heavy-ion collisions has been investigated using a number of
different theoretical approaches \cite{Klein:2003vd,Goncalves:2005yr,
	AyalaFilho:2008zr,
	Cisek:2012yt,Lappi:2013am,
	Guzey:2013xba,Santos:2014zna,Goncalves:2017wgg,Guzey:2016piu,Xie:2016ino,Kopp:2018xvu}. 
In this case the heavy quark mass provides a hard
scale which ensures a dominant contribution from short distances, so that
a perturbative QCD approach becomes applicable. The diffractive photoproduction
then becomes a sensitive probe of the gluon structure of the target.

Much attention has been paid in the past on diffractive photo- and electroproduction
of vector mesons on the proton. A large body of data has been accumulated at 
the DESY-HERA facility. For a review of experimental data and of the theoretical
approaches, see \cite{Ivanov:2004ax}.
Here we will use the color-dipole approach, which allows us to take into account
nuclear effects once the dipole cross section on a free nucleon has been fixed.
To this end we take advantage of available data on exclusive $J/\psi$ production
to check a variety of dipole cross sections against them.
In view of the later application to ultraperipheral heavy-ion collisions the
HERA energy range is the most relevant to us.

Here we discuss the coherent diffractive photoproduction in the same approach which
we used earlier for the incoherent photoproduction of $J/\psi$ \cite{Luszczak:2017dwf}.
This work is organized as follows: in Section \ref{sec:formalism} we review the 
formalism and main formulas for diffractive vector meson production on nucleons
and nuclei in the color dipole approach. In Section \ref{section:dipole_models}
we review different parametrizations/fits of the dipole cross section.
Then, in Section \ref{section:numerical_results} we compare the results of our
numerical results to available published and preliminary experimental data.
We summarize our findings in Section \ref{section:summary}.

\section{Coherent photoproduction in the color dipole approach}
\label{sec:formalism}
\subsection{Free nucleon target}
Let us start with a brief review of the formalism for production
of a vector mesons $V$ of mass $M_V$ (in this work we concentrate on  $V = J/\psi$) at high enough energies,
so that the coherence length $l_c = 2 \omega /M_V^2$ is much larger 
than the size of the proton $l_c \gg R_N$, where $\omega$ is the photon energy. 
In such a situtation the $J/\psi$ photoproduction can be described as a 
elastic scattering of a $c \bar c$ of size $r$ conserved during the interaction (see e.g. \cite{Nikolaev:1992si}). The $\gamma \to c \bar c$ transition and projection of the $c \bar c$ pair on the bound state are 
encoded in the relevant light-cone wave functions, which depend also on the fraction $z$ of the photon's light-front momentum
carried by the quark.
The coherent diffractive amplitude on the free nucleon then takes a form
\begin{eqnarray}
\mathcal{A}(\gamma N \rightarrow V N ;W,\bq)  &=& 2 (i + \rho_N) \int d^2\bb \exp[i \bb \bq] \bra{V} 
\exp[i(1-2z) \br \bq/2] \Gamma_N(x,\bb,\br) \ket{\gamma} \nonumber \\
&=&  (i+\rho_N) \int d^2\br \, \rho_{V \leftarrow \gamma}(\br,\bq)  \sigma(x,\br,\bq) \nonumber \\
&\approx&  (i + \rho_N) \int d^2\br \, \rho_{V \leftarrow \gamma}(\br,0)  \sigma(x,r) \, \exp[-B\bq^2/2] \, . 
\label{eq:amplitude_N}
\end{eqnarray}
Here $x=M_V^2/W^2$, where $W$ is the $\gamma p$-cms energy.
Our amplitude is normalized such that the differential cross section is obtained from
\begin{eqnarray}
{d \sigma( \gamma N \to V N;W) \over dt} = {d \sigma( \gamma N \to V N;W) \over d\bq^2} = {1 \over 16 \pi} 
\Big| \mathcal{A}(\gamma^* N \rightarrow V N ;W,\bq)\Big|^2 \, .
\end{eqnarray}
The overlap of light-front wave functions of photon and the vector meson is
\begin{eqnarray}
\rho_{V \leftarrow \gamma}(\br, \bq) = \int_0^1 dz \Psi_V(z,\br) \Psi_\gamma(z,\br) \exp[i(1-2z)\br\bq/2] \, .
\end{eqnarray}
Here a sum over quark and antiquark helicities is implicit. The overlap depends also on photon and vector meson
helicities and in general gives rise to nonzero helicity flip transitions. We concentrate on the helicity
conserving amplitude, as helicity flip transitions for heavy vector mesons are suppressed. 
In this case one can also neglect the $\bq$-dependent phase factor. 
\footnote{An form of the phase which doesn't vanish at $z = 1/2$ is common in the literature. Note that our phase is consistent with the Feynman-diagram calculations in momentum space \cite{Ivanov:2004ax}.}    
All the dependence on transverse momentum transfer $\bq$ is contained then in the off-forward dipole cross
section, for which we assume a factorized form
\begin{eqnarray}
 \sigma(x,\br,\bq) = \sigma(x,r) \exp[-B\bq^2/2] \, .
\end{eqnarray}
The diffraction slope $B$ depends on energy, for the explicit parametrization used,
see section \ref{Section:Results_proton} below.
Explicitly, the overlap of vector meson and photon light-cone wave function, 
obtained from the $\gamma_\mu$-vertex for the $Q \bar Q \to V$ vertex is given by \cite{Nemchik:1994fp,Nemchik:1996cw}
\begin{eqnarray}
\Psi^*_V(z,\br) \Psi_\gamma(z,\br) &=& { e_Q \sqrt{4 \pi \alpha_{\rm em} } N_c \over 4 \pi^2 z(1-z)} \Big\{ m_Q^2 K_0(m_Q r) \psi(z,r) 
\nonumber \\
&&- [z^2 + (1-z)^2] m_QK_1(m_Q r) 
{\partial \psi(z,r) \over \partial r}   \Big\} .
\end{eqnarray}
For the radial wave function $\psi(z,r)$, we choose the so-called ``boosted Gaussian'' wave function 
\cite{Nemchik:1994fp,Nemchik:1996cw} as parametrized in 
\cite{Kowalski:2006hc} for the $J/\psi$ meson. 

The real part of the amplitude is restored from analyticity from the
$x$-dependent effective intercept
\begin{eqnarray}
\Delta_\Pom = {\partial \log \Big( \bra{V} \sigma(x,r) \ket{\gamma} \Big) \over \partial \log(1/x)} \, , 
\end{eqnarray}
so that 
\begin{eqnarray}
\rho_N = \tan\Big( {\pi \Delta_\Pom \over 2} \Big) \, .
\end{eqnarray}
As we discuss below, the color dipole cross section has been obtained from a fit of the total photoabsorption cross
section on the nucleon, i.e. a fit to the absorptive part of the forward, (Mandelstam $t=0$),  Compton amplitude. 
In vector meson production, even at $\bDelta =0$ the $t=0$ limit is not reached at finite energy and
there is always a finite $t_{\rm min}$ due to the vector meson mass. Consequently, gluons exchanged in the amplitude
carry different longitudinal momenta, at small $x=M_V^2/W^2$ we have typically, say $x_1 \sim x, x_2 \ll x_1$. In such a situation,
the corresponding correction which multiplies the amplitude is Shuvaev's \cite{Shuvaev:1999ce} factor
\begin{eqnarray}
R_{\rm skewed} = { 2^{2 \Delta_\Pom + 3} \over \sqrt{\pi} } \cdot {\Gamma(\Delta_\Pom + 5/2) \over \Gamma(\Delta_\Pom+ 4)} \, .
\end{eqnarray}
This correction has been studied with some rigour only for the two-gluon ladder, where it accounts for the ``skewedness''
of gluon momentum fractions. It is to be applied only at small-$x$.

\subsection{Nuclear Target}

\begin{figure}[!h]
	\begin{center}
		\includegraphics[width=.3\textwidth,angle = 270]{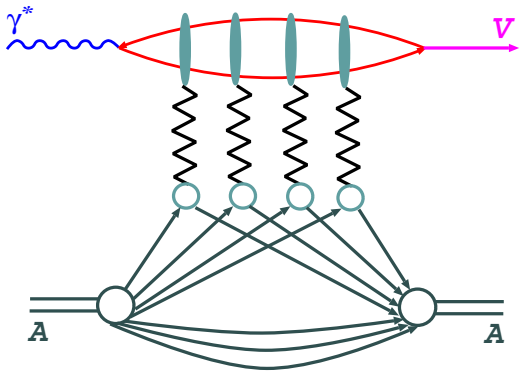}
		\caption{Coherent photoproduction of a vector meson in which the nucleus stays in its ground state.  
		}
		\label{fig:diagrams}
	\end{center}
\end{figure}

When it comes to nuclear targets one should realise that color dipoles can be regarded as eigenstates
of the interaction and one can apply the standard rules of Glauber theory \cite{Glauber} for each 
of the eigenstates. 
We now require that the coherence length be much larger than the nuclear size, $l_c \gg R_A$.
Then we can obtain the Glauber form of the dipole scattering amplitude:
\begin{eqnarray}
\Gamma_A(x,\bb,\br) = 1 - \exp[-\half \sigma(x,r) T_A(\bb)] \, .
\label{eq:Glauber}
\end{eqnarray}
Notice that being states of fixed size color dipoles are not eigenstates of a mass operator, and
the rescattering of dipoles (see Fig.\ref{fig:diagrams}) corresponds to the diffractive transitions $M_i^2 \to M_j^2$
in the individual scatterings. The dipole rescattering therefore is a particular realization of
Gribov's generalization of Glauber theory \cite{Wilkin}. The inelastic shadowing corrections
will include masses which are different from $M_V^2$, but not much larger. That is
to say that inelastic shadowing corresponding to explicit triple Pomeron terms are not included.
  
The dipole amplitude of eq.{\ref{eq:Glauber} corresponds to a rescattering 
of the dipole in a purely absorptive medium. The real part of the dipole-nucleon amplitude 
is often neglected. 
It induces the refractive effects \cite{Glauber}, and instead of eq.{\ref{eq:Glauber} we should take
\begin{eqnarray}
\Gamma_A(x,\bb,\br) &=& 1 - \exp[-\half \sigma(x,r)(1-i \rho_N) T_A(\bb)] \nonumber \\ 
&=& 
1 - \exp[-\half \sigma(x,r) T_A(\bb)] \cos\Big(\half \rho_N \sigma(x,r) T_A(\bb)\Big) \nonumber \\
&& + i \exp[-\half \sigma(x,r) T_A(\bb)] \sin\Big(\half \rho_N \sigma(x,r) T_A(\bb) \Big) \, . 
\end{eqnarray}
Note that the real part of the dipole amplitude $\Gamma_A$ contributes to the imaginary (''absorptive'') part
of the diffractive amplitude, while the imaginary part of $\Gamma_A$ yields the real (''dispersive'') part
of the diffractive amplitude.
We adopt the standard assumption of the nucleus being a dilute gas of uncorrelated nucleons. 
The optical thickness $T_A(\bb)$ is calculated from a Wood-Saxon
distribution $n_A(\vec{r})$:
\begin{eqnarray}
T_A(\bb) = \int_{-\infty}^{\infty} dz \, n_A(\vec{r})  \,;\, \vec{r}=(\bb,z), \, \int d^2\bb \, T_A(\bb) = A. 
\end{eqnarray}
The diffractive amplitude in $\bb$-space is
\begin{eqnarray}
\mathcal{A}(\gamma A \rightarrow V A ;W,\bb)  &=& 2i  \, \bra{V}\Gamma_A(x,\bb,\br) \ket{\gamma} \, {\cal F}_A(q_z). 
\end{eqnarray}
We denote by $W$ the per-nucleon cms-energy in the $\gamma A$-collision.
The nuclear form factor ${\cal F}_A(q) = \exp[-R^2_{\rm ch} q^2/6]$ depends on the 
finite longitudinal momentum transfer $q_z= x m_N$. It serves to cut-off the 
diffractive contribution at low energies (large $x$) where the coherence condition 
is not satisfied.
The total cross section for the $\gamma A \to V A$ reaction is finally obtained as
\begin{eqnarray}
\sigma(\gamma A \to V A;W) = {1 \over 4} \int d^2 \bb \, \Big| \mathcal{A}(\gamma A \rightarrow V A ;W,\bb)    \Big|^2 \, .
\end{eqnarray}

\section{Dipole models}
\label{section:dipole_models}

In the dipole picture the deep inelastic scattering is viewed as a two stage process; 
first the virtual photon fluctuates into a dipole, which consists of  a quark-antiquark pair 
(or a $q\bar q g$ or $q\bar q gg$ ... system) 
and in the second stage the dipole interacts with the 
proton.
Dipole denotes a quasi-stable quantum mechanical state, 
which has a very long life time ($\approx 1/m_p x\;$) and a size $r$, which remains  
unchanged during scattering. 
The wave function $\Psi^{\gamma^*}_{T,L}(z,\br,Q^2)$ determines the probability amplitude 
to find a dipole of size $r$ within a photon. 
This probability depends  on the value of external $Q^2$ and the fraction of the 
photon momentum carried by the quarks forming the dipole, $z$. 

The scattering amplitude is a product of the virtual photon wave function, $\Psi$, with the dipole cross section, $\sigma(x,r)$, 
which determines a probability of the dipole-proton scattering. 
Thus, within the dipole formulation of the $\gamma^* p$ scattering \cite{Nikolaev:1990ja}
\begin{eqnarray}
\sigma_{T,L}(\gamma^* p;x,Q^2) =  \int d^2\br \int_0^1 dz \Big| \Psi^{\gamma^*}_{T,L}(z,\br,Q^2) \Big|^2 \, \sigma(x,r) \, ,
\end{eqnarray}
where $T,L$ denotes the virtual photon polarization and $\sigma_{T,L}^{\gamma^* p}$ the total inclusive DIS cross section.
It is worth to remember that besides a contribution of dipoles of sizes $r^2 \sim 1/Q^2$, the total photoabsorption
also gets a scaling contribution from large dipole sizes $r^2 \sim 1/m_f^2$, where $m_f$ is a 
mass of the quark of flavour $f$.

Numerous models for the  dipole cross section have been developed to test various aspects of the data
In the following we will shortly review some of them, which have been obtained from fitting data on the
inclusive proton structure function and which we will test against the $J/\psi$ photoproduction data before
we proceed to the calculation of the nuclear observables.   

\subsection{GBW model}

The dipole model became a popular tool in investigations of deep-inelastic scattering 
following the observation of Golec-Biernat and W\"usthoff (GBW) \cite{GolecBiernat:1998js,GolecBiernat:1999qd}, 
that a simple ansatz for the dipole cross section was able to describe 
simultaneously the total inclusive and diffractive cross sections measured at HERA.

In the GBW model the dipole-proton cross section $\sigma_{\text{dip}}$ is given by
\begin{equation}
\label{eGBW}
   \sigma(x,r) = \sigma_{0} \left(1 - \exp \left[-\frac{Q^2_s(x)r^{2}}{4} \right]\right),
\end{equation}
where $Q_s^{2}(x)$ is the $x$ dependent saturation scale.
It is parametrized in the form
\begin{equation}
Q_s^2(x) = Q_0^2 \cdot \Big( {x_0 \over x} \Big)^\lambda \, . 
\end{equation}  
The free fitted parameters are: the cross-section normalisation, $\sigma_{0}$, as well as $x_{0}$ and $\lambda$.  
For dipole sizes which are large in comparison to the saturation radius, $R \sim 1/Q_s$, 
the dipole cross section saturates by approaching a constant value $\sigma_0$, 
i.e. saturation damps the growth of the gluon densitIn this model saturation is taken into account 
in the eikonal approximation and 
the saturation radius can be related to the gluon density in the transverse plane. y at low $x$. 

The GBW model provided a good description of data from medium $Q^2$ values ($\approx 30$ GeV$^2$) 
down to low $Q^2$ ($\approx 0.1$ GeV$^2$).
Despite its success and its appealing  simplicity  the model has some  shortcomings; in particular it describes 
the QCD evolution by a simple $x$ dependence,  $ \sim (1/x)^\lambda$, i.e the $Q^2$ dependence of the cross section 
evolution is solely induced by the saturation effects. 
Therefore, it does not  match with DGLAP QCD evolution, which is known to describe 
data very well from $Q^2 \approx 4$ GeV$^2$ to very large $Q^2 \approx 10000$ GeV$^2$.       
However, we have to remember, that for the case of $J/\psi$ production the hard scale is 
just at the lower range of the perturbative regime: $Q^2 \approx M_{J/\psi}^2 /4 \sim 2.5 \, \rm GeV^2$.
Therefore one may expect that the DGLAP evolution effects are not very strong.

In this work we use a new fit of the GBW-form of the dipole cross section obtained by 
Golec-Biernat and Sapeta in \cite{Golec-Biernat:2017lfv}. We take the parameters which they obtained
by fitting HERA data for $Q^2 < 5 \, \rm{GeV}^2$, and which read
$\sigma_0 = 28.18 \, \rm{mb}$, $\lambda = 0.237$ and $x_0 = 0.31 \cdot 10^{-4}$, 
with $Q_0^2 = 1 \, \rm{GeV}^2$.
\subsection{BGK model}
The evolution ansatz of the GBW model was improved in the model proposed by Bartels, Golec-Biernat and Kowalski, 
(BGK)~\cite{Bartels:2002cj},  by taking into account the  DGLAP evolution of the gluon density in an explicit way. 
The model preserves the GBW eikonal approximation to saturation and thus the dipole cross section is given by
\begin{equation}
\label{eBGK}
   \sigma(x,r) = \sigma_{0} \left(1 - \exp \left[-\frac{\pi^{2} r^{2} \alpha_{s}(\mu^{2}) xg(x,\mu^{2})}{3 \sigma_{0}} \right]\right).
\end{equation}
The evolution scale $\mu^{2}$ is connected to the size of the dipole by $\mu^{2} = C/r^{2}+\mu^{2}_{0}$. 
This assumption allows to treat  consistently the contributions of large   
without making the strong coupling constant, $\alpha_s(\mu^2$), unphysically large.

The gluon density, which  is parametrized  at the starting scale $\mu_{0}^{2}$, 
is evolved to larger scales, $\mu^2$, using LO or NLO DGLAP evolution.
For the initial condition, we consider here the {\it soft} ansatz, as used in the original BGK model 
\begin{equation}
   xg(x,\mu^{2}_{0}) = A_{g} x^{-\lambda_{g}}(1-x)^{C_{g}},
\label{gden-soft}
\end{equation}

The free parameters for this model are $\sigma_{0}$, $\mu^{2}_{0}$ and the  parameters for gluon $A_{g}$, $\lambda_{g}$, $C_{g}$.
Their values have been obtained by a fit to the data using the xFitter framework \cite{xfitter} in Ref.\cite{Luszczak:2016bxd}. 
The fit results were found to be independent on the parameter  $C$, which was therefore fixed as  $C=4$ GeV$^2$, 
in agreement with the original BGK fits.
For convenience we show the parameters in Table \ref{tabl1}.

\begin{table}[htbt]
	\begin{center}
		\begin{tabular}{|c|c|c|c|c|c|c|c|c|c|c|} 
			\hline 
			$Q^2_{min}$ [GeV$^2$] &
			$\sigma_0[{\rm mb}]$ & $A_g$ & $\lambda_g$ & $C_g$ & $N_{df}$& $\chi^2$& $\chi^2/N_{df}$\\
			\hline
			$3.5 $&  89.99$\pm$ &  2.44$\pm$ & -0.079$\pm$ & 7.24$\pm$&  530 &  540.35 &  1.02 \\
			& 9.2 & 0.145&  0.099&  0.61 & & &  \\
			\hline
		\end{tabular}
	\end{center}
	\caption{ BGK fit with fitted valence quarks for $\sigma_r$  for H1ZEUS-NC data in the range $Q^2 \ge 3.5$~GeV$^2$  and $x\le 0.01$. NLO fit. { \it Soft gluon}.   $ m_{uds}= 0.14, m_{c}=1.3$ GeV.  $Q_0^2=1.9$ GeV$^2$.}
	\label{tabl1}
\end{table}

\subsection{IIM model}
Another parametrization of the dipole cross section which gives the latter in a
simple analytic form is the IIM model \cite{Iancu:2003ge}.
It is also meant to take into account the saturation effects. 
While the GBW and BGK models use for saturation the eikonal approximation,  
the IIM model uses a simplified version of the  Balitsky-Kovchegov equation~\cite{Balitsky:1995ub,Kovchegov:1999yj}. 
Here we use a parametrization obtained by Soyez which includes heavy quarks into the fit \cite{Soyez:2007kg}.  
The model was compared  with the recent H1 data in~\cite{H1FL},  
where it was shown that it provides a good data description in the lower $Q^2$ range, $0.2<Q^2< 40$ GeV$^2$. 
As this model also applies to the range of moderately large $Q^2$ it is an appropriate choice 
for our problem.
The dipole cross section is parametrized as
\begin{eqnarray}
\sigma(x,r) = 2 \pi R_p^2 
\begin{cases}
N_0 \exp[-2\gamma L - {L^2 \over \kappa \lambda Y}]  & \text{if}\ L \geq 0, \\
1 - \exp[-a (L-L_0)^2]  & \text{else},
\end{cases}
\end{eqnarray}
where
\begin{eqnarray}
L = \log\Big( {2 \over r Q_s} \Big) \, , \, Q_s^2 =   \Big({x_0 \over x}\Big)^\lambda {\rm GeV}^2, \, Y = \log \Big({1 \over x} \Big) \, .
\end{eqnarray}
and
\begin{eqnarray}
L_0 = {1 - N_0 \over \gamma N_0} \, \log\Big( {1 \over 1 -N_0} \Big) \, , a ={1 \over L_0^2}  \log\Big( {1 \over 1 -N_0}\Big).
\end{eqnarray}
We take the numerical values found in the xFitter code:
\begin{eqnarray}
N_0 = 0.7, R_p = 3.44 \, {\rm GeV}^{-1}, \, \, \gamma = 0.7376, \kappa = 9.9 \, , 
\lambda = 0.2197, \, x_0 = 1.632 \cdot 10^{-5}.
\end{eqnarray}

\section{Numerical results}
\label{section:numerical_results}

\subsection{Predictions for  $J/\psi$ production on the proton target}
\label{Section:Results_proton}

Let us now turn to the numerical results we obtained for the total exclusive 
photoproduction cross section of $J/\psi$ on the proton target. 
For the GBW and IIM dipole cross sections, we calculate the total cross section from 
\begin{eqnarray}
\sigma(\gamma p \to J/\psi p;W) = {1 + \rho_N^2\over 16 \pi B} \, R_{\rm skewed}^2 \, |\bra{V} \sigma(x,r) \ket{\gamma}|^2 \, .
\end{eqnarray}
The diffraction slope $B$ is taken as $B = B_0 + 4 \alpha'\, \log(W/W_0)$, with
$B_0 = 4.88 \, \rm{GeV}^{-2}$, $\alpha' = 0.164 \, \rm{GeV}^{-2}$, and $W_0 = 90 \, \rm{GeV}$.
For the BGK type of parametrizations, it proves to be more stable numerically to 
substitute the ``skewed glue'' in the exponent:
\begin{equation}
\label{eBGK}
\sigma(x,r) = \sigma_{0} \left(1 - \exp \left[-\frac{\pi^{2} r^{2} \alpha_{s}(\mu^{2}) R_{\rm skewed} xg(x,\mu^{2})}{3 \sigma_{0}} \right]\right),
\end{equation}
where the exponent $\Delta_\Pom$ which enters the Shuvaev-factor is calculated from the  
relevant gluon distribution. This avoids taking tedious derivatives in the numerical
grid for the dipole cross section.
\begin{figure}[!h]
	\begin{center}
		\includegraphics[width=.8\textwidth]{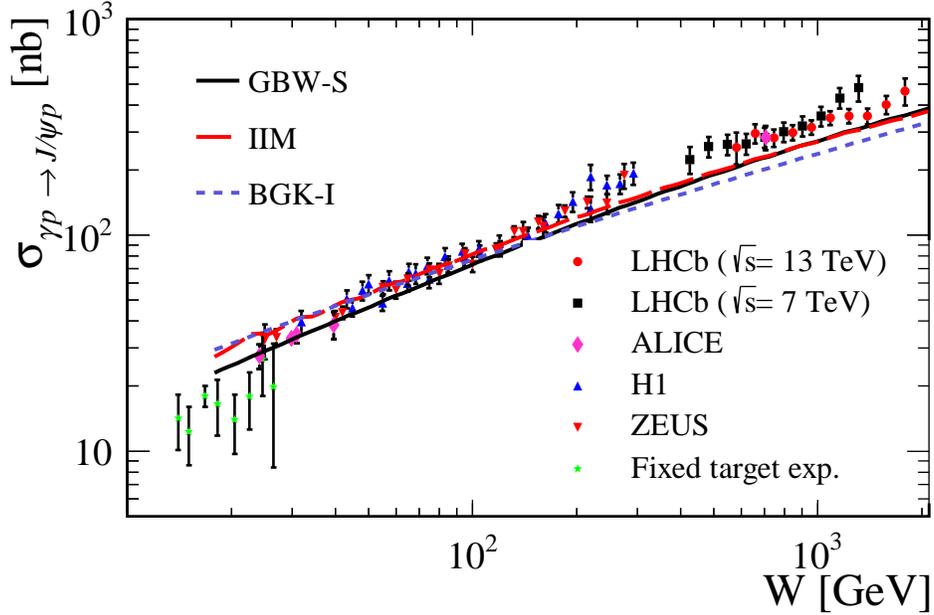}
		\caption{Total cross section for the exclusive photoproduction $\gamma p \to J/\psi p$ as a function of			    
			$\gamma p$-cms energy $W$. The data are from Refs. \cite{Binkley:1981kv,Denby:1983az,Frabetti:1993ux,Alexa:2013xxa,Aktas:2005xu,Chekanov:2002xi,Aaij:2014iea,Aaij:2018arx,TheALICE:2014dwa}.
			The results for three different dipole cross-sections are shown.}
		\label{fig:sig_gamma_p}
	\end{center}
\end{figure}
Our results are shown if Fig.\ref{fig:sig_gamma_p}, where we compare the results from the three different parametrizations
of the dipole cross section against the data from Refs. \cite{Binkley:1981kv,Denby:1983az,Frabetti:1993ux,Alexa:2013xxa,Aktas:2005xu,Chekanov:2002xi,Aaij:2014iea,Aaij:2018arx,TheALICE:2014dwa} We observe that the range of $30 \lsim W \lsim 300 \rm GeV$ is reasonably
well described by all dipole cross sections.  
The very high-energy domain is covered
by data which have been extracted from the $pp \to pp J/\psi$ reaction by the LHCb collaboration \cite{Aaij:2014iea,Aaij:2018arx}.
Here none of the models does a good job. 
While one could certainly try to obtain a dipole cross section which also fits the high-energy data, possibly by including
vector meson data into the fit, this is not necessary for the purpose of this paper.
Namely it is the HERA energy-range which will be most crucial
for the calculations in ultraperipheral heavy-ion collisions later on.
In that respect, the description of free-nucleon data for the chosen dipole cross sections 
is satisfactory for our purposes. 

\subsection{Results for photoproduction in ultraperipheral collisions}

\begin{figure}[!h]
	\begin{center}
		\includegraphics[width=.4\textwidth]{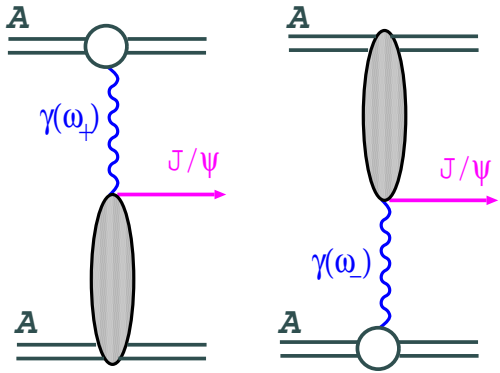}
		\caption{Exclusive photoproduction in ultraperipheral heavy-ion collisions.}
		\label{fig:diagrams_AA}
	\end{center}
\end{figure}

\begin{table}[htbt]
	\begin{center}
		\begin{tabular}{|c|c|c|c|c|c|c|c|c|c|c|c|} 
			\hline
			$\sqrt{s_{NN}} = 2.76 \, \rm TeV$ & & & & & & & & \\
			\hline 
			$y$ &
			$W_+[\rm GeV]$ & $W_-[\rm GeV]$ & $x_+$ & $x_-$ & $n(\omega_+)$& $n(\omega_-)$& $\sigma(W_+ ) [\mu \rm b]$ & $\sigma(W_-) [\mu \rm b]$ \\
			\hline
			0.0 & 92.5 & 92.5 & $1.12 \cdot 10^{-3}$ & $1.12 \cdot 10^{-3}$ & 69.4 & 69.4 & 27.4 &  27.4  \\
			\hline
			1.0 & 152 & 56.1 & $ 4.13 \cdot 10^{-4}$ & $ 3.05 \cdot 10^{-3}$ & 39.5 & 100 & 37.1 &  19.5  \\
			\hline
			2.0 & 251 & 34.0 & $ 1.52 \cdot 10^{-4}$ & $ 8.29 \cdot 10^{-3}$ & 14.5 & 132 & 48.6 &  13.0  \\
			\hline
			3.0 & 414 & 20.6 & $ 5.59 \cdot 10^{-5}$ & $ 2.25 \cdot 10^{-2}$ & 1.68 & 163 & 62.2 &  7.18  \\
			\hline
			3.8 & 618 & 13.8 & $ 2.51 \cdot 10^{-5}$ & $ 5.02 \cdot 10^{-2}$ & 0.03 & 188 & 74.6 &  2.81  \\
			\hline
		\end{tabular}
	\end{center}
	\caption{Subenergies $W_\pm$ and Bjorken-$x$ values $x_\pm$ for $\sqrt{s_{NN}}= 2.76 \, \rm TeV$ for a given rapidity $y$. Also shown are photon fluxes $n(\omega_\pm)$
		and the photoproduction cross sections on $^{208}{\rm Pb}$ at energies $W_\pm$ for the IIM-dipole cross section.}
	\label{table:4}
\end{table}

\begin{table}[htbt]
	\begin{center}
		\begin{tabular}{|c|c|c|c|c|c|c|c|c|c|c|c|} 
			\hline
			$\sqrt{s_{NN}} = 5.02 \, \rm TeV$ & & & & & & & & \\
			\hline 
			$y$ &
			$W_+[\rm GeV]$ & $W_-[\rm GeV]$ & $x_+$ & $x_-$ & $n(\omega_+)$& $n(\omega_-)$& $\sigma(W_+ ) [\mu \rm b]$ & $\sigma(W_-) [\mu \rm b]$ \\
			\hline
			0.0 & 125 & 125 & $6.17 \cdot 10^{-4}$ & $6.17 \cdot 10^{-4}$ & 87.9 & 87.9 & 32.9 &  32.9  \\
			\hline
			1.0 & 206 & 75.6 & $ 2.27 \cdot 10^{-4}$ & $ 1.68 \cdot 10^{-3}$ & 57.2 & 119 & 43.8 &  24.0  \\
			\hline
			2.0 & 339 & 45.9 & $ 8.35 \cdot 10^{-5}$ & $ 4.56 \cdot 10^{-3}$ & 28.5 & 150 & 56.5 &  16.8  \\
			\hline
			3.0 & 559 & 27.8 & $ 3.07 \cdot 10^{-5}$ & $ 1.24 \cdot 10^{-2}$ & 7.5 & 181 & 71.3 &  10.6  \\
			\hline
			4.0 & 921 & 16.9 & $ 1.13 \cdot 10^{-5}$ & $ 3.37 \cdot 10^{-2}$ & 0.35 & 213 & 88.6 &  4.98  \\
			\hline
			4.8 & 1370 & 11.3 & $ 5.08 \cdot 10^{-6}$ & $ 7.50 \cdot 10^{-2}$ & 0.001 & 238 & 103 &  1.22  \\
			\hline
		\end{tabular}
	\end{center}
	\caption{Subenergies $W_\pm$ and Bjorken-$x$ values $x_\pm$ for $\sqrt{s_{NN}}= 5.02 \, \rm TeV$ for a given rapidity $y$. Also shown are photon fluxes $n(\omega_\pm)$
		and the photoproduction cross sections on $^{208}{\rm Pb}$ at energies $W_\pm$ for the IIM-dipole cross section.}
	\label{table:5}
\end{table}

We now turn to our results for ultraperipheral heavy-ion collisions. We obtain the rapidity-dependent
cross section for exclusive $J/\psi$ production from the Weizs\"acker-Williams fluxes of quasi-real photons $n(\omega)$
as
\begin{eqnarray}
{d \sigma (AA \to AA J/\psi;\sqrt{s_{NN}}) \over dy} = 
n(\omega_+) \sigma(\gamma A \to J/\psi A; W_+) + n(\omega_-) \sigma(\gamma A \to J/\psi A; W_-) \, . \nonumber \\
\end{eqnarray}
Here the two terms correspond to the contributions where the left-moving ion serves as the photon source and 
the right-moving one as the target and vice-versa
(see Fig. \ref{fig:diagrams_AA}). Note that we neglected the interference between the two
processes. This interference is concentrated at very small transverse momenta \cite{Klein:2003vd,Hencken:2005hb,Schafer:2007mm}.
It introduces the azimuthal correlation between the outgoing ions, and in the absence of absorptive corrections
it would vanish after the angular integration \cite{Schafer:2007mm}.

We use the standard form of the Weizs\"acker-Williams flux (see e.g. the reviews \cite{Bertulani:2005ru,Baur:2001jj}) for the ion
moving with boost $\gamma$: 
\begin{eqnarray}
n(\omega) = {2 Z^2 \alpha_{\rm em} \over \pi} \Big[ \xi K_0(\xi) K_1(\xi) - {\xi^2 \over 2} (K_1^2(\xi) -K_0^2(\xi))  \Big].
\end{eqnarray}
Here $\omega$ is the photon energy, and $\xi = 2R_A \omega/\gamma$. This flux was obtained by imposing the constraint
on the impact parameter of the collision $b > 2R_A$, where we use $R_A= 7 \,\rm{fm}$. 
This means that configurations where nuclei touch each other are excluded, as otherwise inelastic processes would destroy
the rapidity gaps in the event.
The photon energies corresponding to the two contributions are
$\omega_\pm = m_V \exp[\pm y]/2$, the corresponding cms-energies for the $\gamma A \to J/\psi A$ subprocesses
are $W_\pm = 2 \sqrt{s_{NN}} \omega_\pm$.

In order to understand the kinematics a bit better,
in Tables \ref{table:4} and \ref{table:5} we show the values of $W_\pm$, the associated 
Bjorken-$x$ values $x_\pm$ as well as photon fluxes $n(\omega_\pm)$. 
For convenience we have also included the values of the photoproduction cross section $\sigma(\gamma A \to J/\psi A:W_\pm)$
for the example of the IIM dipole cross section, and for the $^{208} \rm Pb$ nucleus.
Table \ref{table:4} is for the energy $\sqrt{s_{NN}} = 2.76 \, \rm TeV$ and Table \ref{table:5} for  
$\sqrt{s_{NN}} = 5.02 \, \rm TeV$. At midrapidity of course $W_\pm$ coincide, and we are always well in the 
energy range that has been explored at HERA for the free nucleon target.
If we move out to larger rapidities of the two processes it is the {\it low-energy} reaction which tends to dominate.
This has its explanation in the rather quick drop of the nuclear photon fluxes at high photon energies and the modest
rise of the nuclear photoproduction cross section.

In figure \ref{fig:dsig_dy_2760} we show the cross section as a function of $J/\psi$-rapidity, for $\sqrt{s_{NN}} = 2.76 \, \rm TeV$
for the three different dipole cross sections introduced previously.
We compare them to the data obtained by the ALICE \cite{Abelev:2012ba,Abbas:2013oua} and CMS \cite{Khachatryan:2016qhq} 
collaborations. We see that at large rapidities we obtain a fair description of the data, while all of the dipole
cross sections overestimate the data at mid-rapidity.

In figure \ref{fig:dsig_dy_5020_ALICE} the rapidity dependent cross section is shown at $\sqrt{s_{NN}} = 5.02 \, \rm TeV$
and compared to preliminary data from the ALICE collaboration \cite{Kryshen:2017jfz}.
Figure \ref{fig:dsig_dy_5020_LHCb} also is calculated at $\sqrt{s_{NN}} = 5.02 \, \rm TeV$ but compares to the preliminary
data by the LHCb collaboration \cite{LHCb:2018ofh}. We see that again we get a reasonable desription of
the preliminary data.
We should point out that we did not include a skewedness
correction to our results on nuclear targets.
It is not completely clear how the skewedness correction
should be applied in the case of nuclear rescatterings.
Obviously the assumption of a two-gluon exchange does not apply to our mechanism, and the necessary longitudinal momentum transfer can be shared by many gluons. As the diffractive amplitude has a ``form-factor''-like (perhaps exponential) behaviour an equipartition of longitudinal momentum transfers looks more likely than having one
``large'' momentum transfer and many small ones.
In addition, the data at large rapidities have a large
contribution from not very small $x$ where the 
skewedness correction may not be justified. 
We therefore follow the authors of \cite{Goncalves:2017wgg}
and omit the skewedness correction.

The overall picture suggests that the Glauber-Gribov formalism in the color dipole approach 
works reasonably well at not too high energies (or not too small $x$), while at higher energies (smaller $x$)
there is room for additional nuclear suppression. 
In fact in our calculations we included only the rescattering of the $c \bar c$ pair.
Due to the ``scanning-radius'' property of vector meson photoproduction, in rescatterings 
of the small $c \bar c$ pair are in fact higher twist effects. This is different than
the case of the inclusive structure function $F_2$, where large dipoles contribute to
a scaling (up to logarithms) nuclear shadowing \cite{Nikolaev:1990ja}.

It is well understood that at small-$x$ the coherency condition can be also satisfied by higher Fock 
$c \bar c g$, $c \bar c g g \dots$ states. In fact these Fock states (in a configuration of strongly ordered 
transverse sizes) are responsible for the DGLAP evolution of structure functions, while configurations
strongly ordered in gluon longitudinal momenta correspond to the BFKL/BK limit.
We believe that the missing higher Fock-states are the main culprit behind the overprediction of the ALICE 
data at midrapidity. Whether their effect can eventually be absorbed into a leading twist shadowing correction
to the DGLAP evolving nuclear glue is an open issue at the moment.

One may ask finally, if the light-cone wave function can be the scapegoat. Indeed in a careful analysis of
some theoretical uncertainties \cite{Santos:2014zna} (see also e.g. \cite{Xie:2016ino}) it was shown that
there can be a sizeable dependence on the meson light-cone wave-function. However here we take the point of view 
that the succesful description of free nucleon HERA data in the important for us energy range ``fixes''
the wave-function overlap. Indeed with the dipole cross sections we employed we would rather spoil the
agreement with HERA data if we substitute another of the popular parametrizations.
From this point of view attempts to obtain light cone wave functions from other sources \cite{Chen:2016dlk}
are interesting.

\begin{figure}[!h]
	\begin{center}
		\includegraphics[width=.5\textwidth]{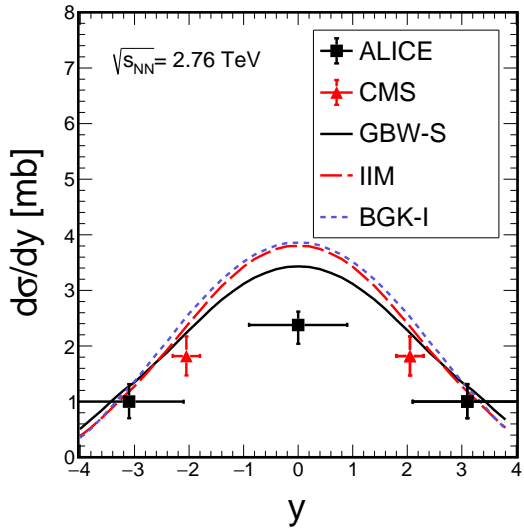}
		\caption{Rapidity dependent cross section $d\sigma/dy$ for exclusive production of $J/\psi$ in $^{208}{\rm Pb} ^{208}{\rm Pb}$-collisions at
			per-nucleon cms energy $\sqrt{s_{NN}} = 2.76 \, \rm TeV$. The data are from ALICE \cite{Abbas:2013oua,Abelev:2012ba} and CMS \cite{Khachatryan:2016qhq}.}
		\label{fig:dsig_dy_2760}
	\end{center}
\end{figure}
\begin{figure}[!h]
	\begin{center}
		\includegraphics[width=.6\textwidth]{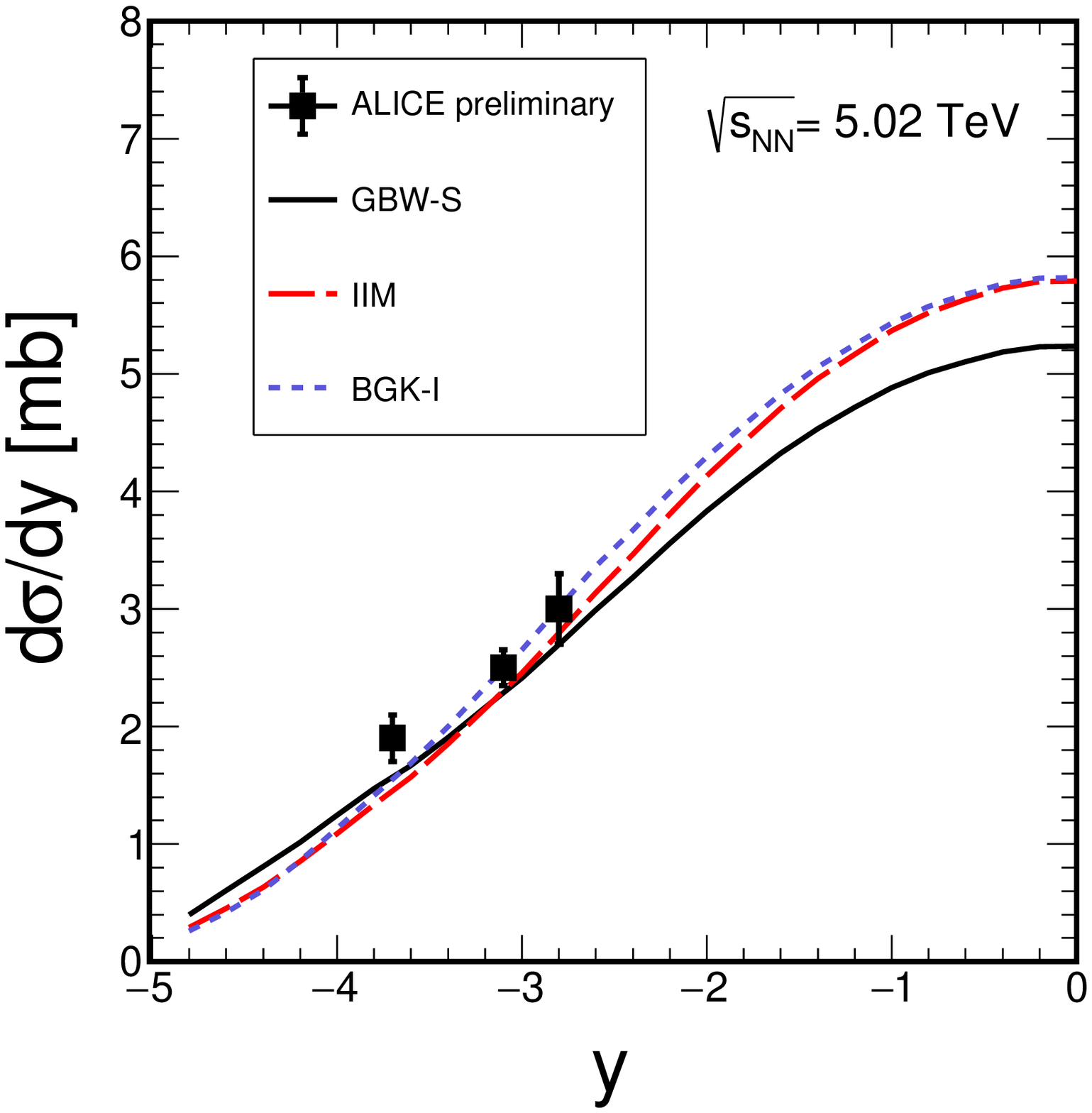}
		\caption{Rapidity dependent cross section $d\sigma/dy$ for exclusive production of $J/\psi$ in $^{208}{\rm Pb} ^{208}{\rm Pb}$-collisions at
			per-nucleon cms energy $\sqrt{s_{NN}} = 5.02 \, \rm TeV$. Shown are preliminary data are from ALICE 
			\cite{Kryshen:2017jfz}.}
		\label{fig:dsig_dy_5020_ALICE}
	\end{center}
\end{figure}
\begin{figure}[!h]
	\begin{center}
		\includegraphics[width=.6\textwidth]{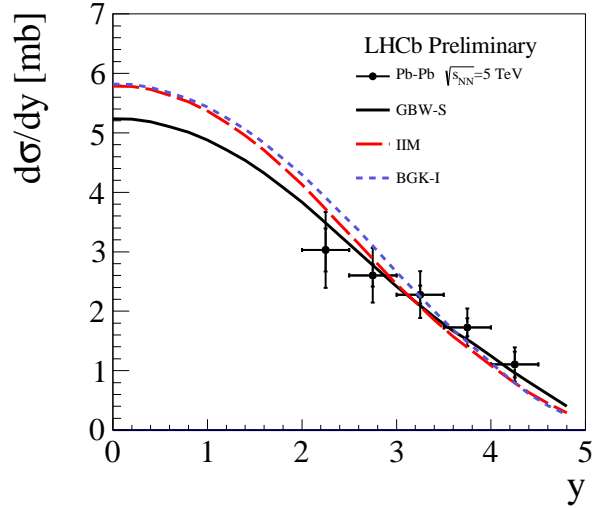}
		\caption{Rapidity dependent cross section $d\sigma/dy$ for exclusive production of $J/\psi$ in $^{208}{\rm Pb} ^{208}{\rm Pb}$-collisions at
			per-nucleon cms energy $\sqrt{s_{NN}} = 5.02 \, \rm TeV$. Shown are preliminary data are from LHCb
			\cite{LHCb:2018ofh}.}
		\label{fig:dsig_dy_5020_LHCb}
	\end{center}
\end{figure}

\section{Summary and outlook}
\label{section:summary}

In this paper we have presented calculations using the Glauber-Gribov theory for coherent 
exclusive photoproduction of $J/\psi$-mesons on heavy nuclei within the color dipole approach.
The dipole cross sections which we utilize have all been obtained from fitting inclusive
deep-inelastic structure function data from HERA.
We first calculated the total elastic photoproduction of $J/\psi$ on the free nucleon
comparing to the data available from fixed-target epxeriments, from the H1 and ZEUS collaborations at 
HERA as well as to data extracted from $pp$ or $pA$ collisions by the LHCB and ALICE collaborations.
All the three dipole cross sections used in this work give a reasonable description of the data, up
to and including the HERA energy range, when
used together with the so-called ``boosted Gaussian'' parametrization of the $J/\psi$ wave function.
The higher energy data extracted mainly by the LHCb collaboration from exclusive $pp$ collisions are 
not well described. 
 
We have applied our results to the exclusive $J/\psi$ production in heavy-ion (lead-lead)
collisions at the energies $\sqrt{s_{NN}}=2.76 \, \rm GeV$ and $\sqrt{s_{NN}} = 5.02 \, \rm GeV$,

The color dipoles play the role of the eigenstates of the scattering
matrix and take into account the  inelastic shadowing corrections. 
We have taken into account the rescattering of a $c \bar c$ dipole in the nucleus taking
into account the real part of the free nucleon amplitude consistent with the
rules of Glauber theory.

Although there is substantial uncertainty as to how to include the skewedness correction
in to the nuclear amplitude, the description of published and preliminary data can be
regarded satisfactory. However the data point taken by ALICE at midrapidity for $\sqrt{s_{NN}}=2.76 \, \rm{TeV}$
is overpredicted. This seems to point to the fact that rescattering of the $c \bar c$ dipole
is insufficient at energies $W_{\gamma A} \sim 100 \, \rm{GeV}$ or $x \sim 0.001$. 

We believe that explicit account of higher Fock-states is necessary in this kinematic region.
This is consistent 
with an analysis of nuclear shadowing and 
deep inelastic diffraction on nuclei in \cite{Nikolaev:2006mh}. There it is shown, that 
$q \bar q g$ states are important for nuclear shadowing
at $x \lsim 0.005$. 

Whether the correct approach is a resummation of their effect in a BFKL/BK framework, or
whether they can be absorbed into a leading twist shadowing of the collinear nuclear glue
is an open issue. It stands to reason that this issue can hardly be resolved by only one
observable, measured essentially at one hard scale, and that future measurements at an electron-ion collider will be crucial for a deeper understanding of the nuclear glue.

\section*{Acknowledgements}

This work is partially supported  by the Polish National Science Center
grant DEC-2014/15/B/ST2/02528. 



\begin{thebibliography}{100}

\bibitem{Klein:1999qj} 
S.~Klein and J.~Nystrand,
Phys.\ Rev.\ C {\bf 60}, 014903 (1999)
[hep-ph/9902259].

\bibitem{Abelev:2012ba} 
B.~Abelev {\it et al.} [ALICE Collaboration],
Phys.\ Lett.\ B {\bf 718}, 1273 (2013)
[arXiv:1209.3715 [nucl-ex]].

\bibitem{Abbas:2013oua}
E.~Abbas {\it et al.} [ALICE Collaboration],
Eur.\ Phys.\ J.\ C {\bf 73} (2013) no.11,  2617
[arXiv:1305.1467 [nucl-ex]].

\bibitem{Khachatryan:2016qhq} 
V.~Khachatryan {\it et al.} [CMS Collaboration],
Phys.\ Lett.\ B {\bf 772}, 489 (2017)
[arXiv:1605.06966 [nucl-ex]].

\bibitem{Kryshen:2017jfz} 
E.~L.~Kryshen [ALICE Collaboration],
Nucl.\ Phys.\ A {\bf 967}, 273 (2017)
[arXiv:1705.06872 [nucl-ex]].

\bibitem{LHCb:2018ofh} 
[LHCb Collaboration],
``Study of coherent $J/\psi$ production in lead-lead collisions at 
$\sqrt{s_{\rm NN}} =5\ \rm{TeV}$ with the LHCb experiment,''
LHCb-CONF-2018-003, CERN-LHCb-CONF-2018-003.

\bibitem{Contreras:2015dqa}
J.~G.~Contreras and J.~D.~Tapia Takaki,
Int.\ J.\ Mod.\ Phys.\ A {\bf 30} (2015) 1542012.

\bibitem{Klein:2003vd} 
S.~R.~Klein and J.~Nystrand,
Phys.\ Rev.\ Lett.\  {\bf 92}, 142003 (2004)
[hep-ph/0311164].

\bibitem{Goncalves:2005yr} 
V.~P.~Goncalves and M.~V.~T.~Machado,
Eur.\ Phys.\ J.\ C {\bf 40}, 519 (2005)
[hep-ph/0501099].

\bibitem{AyalaFilho:2008zr} 
A.~L.~Ayala Filho, V.~P.~Goncalves and M.~T.~Griep,
Phys.\ Rev.\ C {\bf 78}, 044904 (2008)
[arXiv:0808.0366 [hep-ph]].

\bibitem{Cisek:2012yt} 
A.~Cisek, W.~Sch\"afer and A.~Szczurek,
Phys.\ Rev.\ C {\bf 86}, 014905 (2012)
[arXiv:1204.5381 [hep-ph]].

\bibitem{Lappi:2013am} 
T.~Lappi and H.~Mantysaari,
Phys.\ Rev.\ C {\bf 87}, no. 3, 032201 (2013)
[arXiv:1301.4095 [hep-ph]].

\bibitem{Guzey:2013xba} 
V.~Guzey, E.~Kryshen, M.~Strikman and M.~Zhalov,
Phys.\ Lett.\ B {\bf 726}, 290 (2013)
[arXiv:1305.1724 [hep-ph]].

\bibitem{Santos:2014zna} 
G.~Sampaio dos Santos and M.~V.~T.~Machado,
J.\ Phys.\ G {\bf 42}, no. 10, 105001 (2015)
[arXiv:1411.7918 [hep-ph]].

\bibitem{Goncalves:2017wgg} 
V.~P.~Gonçalves, M.~V.~T.~Machado, B.~D.~Moreira, F.~S.~Navarra and G.~S.~dos Santos,
Phys.\ Rev.\ D {\bf 96}, no. 9, 094027 (2017)
[arXiv:1710.10070 [hep-ph]].

\bibitem{Guzey:2016piu} 
V.~Guzey, E.~Kryshen and M.~Zhalov,
Phys.\ Rev.\ C {\bf 93}, no. 5, 055206 (2016)
[arXiv:1602.01456 [nucl-th]].

\bibitem{Xie:2016ino} 
Y.~p.~Xie and X.~Chen,
Eur.\ Phys.\ J.\ C {\bf 76}, no. 6, 316 (2016)
[arXiv:1602.00937 [hep-ph]].

\bibitem{Kopp:2018xvu} 
F.~Kopp and M.~V.~T.~Machado,
Phys.\ Rev.\ D {\bf 98}, 014010 (2018)
[arXiv:1806.06701 [hep-ph]].

\bibitem{Ivanov:2004ax} 
I.~P.~Ivanov, N.~N.~Nikolaev and A.~A.~Savin,
Phys.\ Part.\ Nucl.\  {\bf 37}, 1 (2006)
[hep-ph/0501034].

\bibitem{Luszczak:2017dwf} 
A.~{\L}uszczak and W.~Sch\"afer,
Phys.\ Rev.\ C {\bf 97}, no. 2, 024903 (2018)
[arXiv:1712.04502 [hep-ph]].

\bibitem{Nikolaev:1992si} 
N.~N.~Nikolaev,
Comments Nucl.\ Part.\ Phys.\  {\bf 21}, no. 1, 41 (1992).

\bibitem{Nemchik:1994fp}
J.~Nemchik, N.~N.~Nikolaev and B.~G.~Zakharov,
Phys.\ Lett.\ B {\bf 341} (1994) 228
[hep-ph/9405355].

\bibitem{Nemchik:1996cw}
J.~Nemchik, N.~N.~Nikolaev, E.~Predazzi and B.~G.~Zakharov,
Z.\ Phys.\ C {\bf 75} (1997) 71
[hep-ph/9605231].

\bibitem{Kowalski:2006hc}
H.~Kowalski, L.~Motyka and G.~Watt,
Phys.\ Rev.\ D {\bf 74} (2006) 074016
[hep-ph/0606272].

\bibitem{Shuvaev:1999ce}
A.~G.~Shuvaev, K.~J.~Golec-Biernat, A.~D.~Martin and M.~G.~Ryskin,
Phys.\ Rev.\ D {\bf 60} (1999) 014015
[hep-ph/9902410].


\bibitem{Glauber}	
R.~J.~ Glauber, 
Lectures in Theoretical Physics; W.E. Brittin, L.J. Dunham 
(eds.) pp. 315-414, v. 1, New York: Interscience 1959 

\bibitem{Wilkin}
E.~S.~Abers, H.~Burkhardt, V.~L.~Teplitz, and C.~Wilkin, Nuovo Cimento {\bf 42}, 365 (1966).
J.~Pumplin and M.~H.~Ross,
Phys.\ Rev.\ Lett.\  {\bf 21} (1968) 1778.

V.~N.~Gribov,
Sov.\ Phys.\ JETP {\bf 29} (1969) 483
[Zh.\ Eksp.\ Teor.\ Fiz.\  {\bf 56} (1969) 892].

V.~N.~Gribov,
Sov.\ Phys.\ JETP {\bf 30} (1970) 709
[Zh.\ Eksp.\ Teor.\ Fiz.\  {\bf 57} (1969) 1306].

\bibitem{Nikolaev:1990ja}
N.~N.~Nikolaev and B.~G.~Zakharov,
Z.\ Phys.\ C {\bf 49} (1991) 607.

\bibitem{GolecBiernat:1998js}
K.~J.~Golec-Biernat and M.~W\"usthoff,
Phys.\ Rev.\ D {\bf 59} (1998) 014017
[hep-ph/9807513].

\bibitem{GolecBiernat:1999qd} 
K.~J.~Golec-Biernat and M.~Wusthoff,
Phys.\ Rev.\ D {\bf 60}, 114023 (1999)
[hep-ph/9903358].

\bibitem{Golec-Biernat:2017lfv} 
K.~Golec-Biernat and S.~Sapeta,
JHEP {\bf 1803}, 102 (2018)
[arXiv:1711.11360 [hep-ph]].


\bibitem{Bartels:2002cj} 
J.~Bartels, K.~J.~Golec-Biernat and H.~Kowalski,
Phys.\ Rev.\ D {\bf 66}, 014001 (2002)
[hep-ph/0203258].


\bibitem{xfitter}
S.~Alekhin {\it et al.},
Eur.\ Phys.\ J.\ C {\bf 75} (2015) no.7,  304
[arXiv:1410.4412 [hep-ph]];
H.~Abramowicz {\it et al.} [H1 and ZEUS Collaborations],
Eur.\ Phys.\ J.\ C {\bf 75} (2015) no.12,  580
[arXiv:1506.06042 [hep-ex]];
M.~Botje,
Comput.\ Phys.\ Commun.\  {\bf 182} (2011) 490
[arXiv:1005.1481 [hep-ph]];
F.~James and M.~Roos,
Comput.\ Phys.\ Commun.\  {\bf 10} (1975) 343;
R.~S.~Thorne and R.~G.~Roberts,
Phys.\ Rev.\ D {\bf 57} (1998) 6871
[hep-ph/9709442];
R.~S.~Thorne,
Phys.\ Rev.\ D {\bf 73} (2006) 054019
[hep-ph/0601245];
F.~D.~Aaron {\it et al.} [H1 Collaboration],
Eur.\ Phys.\ J.\ C {\bf 63} (2009) 625
[arXiv:0904.0929 [hep-ex]];
S.~Chekanov {\it et al.} [ZEUS Collaboration],
Eur.\ Phys.\ J.\ C {\bf 42} (2005) 1
[hep-ph/0503274];
J.~Pumplin, D.~R.~Stump and W.~K.~Tung,
Phys.\ Rev.\ D {\bf 65} (2001) 014011
[hep-ph/0008191];
J.~Pumplin, D.~R.~Stump, J.~Huston, H.~L.~Lai, P.~M.~Nadolsky and W.~K.~Tung,
JHEP {\bf 0207} (2002) 012
[hep-ph/0201195].

\bibitem{Luszczak:2013rxa}
A.~{\L}uszczak and H.~Kowalski,
Phys.\ Rev.\ D {\bf 89} (2014) no.7,  074051
[arXiv:1312.4060 [hep-ph]].

\bibitem{Luszczak:2016bxd}
A.~{\L}uszczak and H.~Kowalski,
Phys.\ Rev.\ D {\bf 95} (2017) no.1,  014030
[arXiv:1611.10100 [hep-ph]].



\bibitem{Iancu:2003ge} 
E.~Iancu, K.~Itakura and S.~Munier,
Phys.\ Lett.\ B {\bf 590}, 199 (2004)
[hep-ph/0310338].

\bibitem{Balitsky:1995ub} 
I.~Balitsky,
Nucl.\ Phys.\ B {\bf 463}, 99 (1996)
[hep-ph/9509348].

\bibitem{Kovchegov:1999yj} 
Y.~V.~Kovchegov,
Phys.\ Rev.\ D {\bf 60}, 034008 (1999)
[hep-ph/9901281].

\bibitem{Soyez:2007kg} 
G.~Soyez,
Phys.\ Lett.\ B {\bf 655}, 32 (2007)
[arXiv:0705.3672 [hep-ph]].

\bibitem{H1FL}
F.~D.~Aaron {\it et al.} [H1 Collaboration],
Eur.\ Phys.\ J.\ C {\bf 71}, 1579 (2011)
[arXiv:1012.4355 [hep-ex]].


\bibitem{Binkley:1981kv} 
M.~E.~Binkley {\it et al.},
Phys.\ Rev.\ Lett.\  {\bf 48}, 73 (1982).

\bibitem{Denby:1983az} 
B.~H.~Denby {\it et al.},
Phys.\ Rev.\ Lett.\  {\bf 52}, 795 (1984).

\bibitem{Frabetti:1993ux} 
P.~L.~Frabetti {\it et al.} [E687 Collaboration],
Phys.\ Lett.\ B {\bf 316}, 197 (1993).

\bibitem{Alexa:2013xxa} 
C.~Alexa {\it et al.} [H1 Collaboration],
Eur.\ Phys.\ J.\ C {\bf 73}, no. 6, 2466 (2013)
[arXiv:1304.5162 [hep-ex]].

\bibitem{Aktas:2005xu} 
A.~Aktas {\it et al.} [H1 Collaboration],
Eur.\ Phys.\ J.\ C {\bf 46}, 585 (2006)
[hep-ex/0510016].

\bibitem{Chekanov:2002xi} 
S.~Chekanov {\it et al.} [ZEUS Collaboration],
Eur.\ Phys.\ J.\ C {\bf 24}, 345 (2002)
[hep-ex/0201043].

\bibitem{Aaij:2014iea} 
R.~Aaij {\it et al.} [LHCb Collaboration],
J.\ Phys.\ G {\bf 41}, 055002 (2014)
[arXiv:1401.3288 [hep-ex]].

\bibitem{Aaij:2018arx} 
R.~Aaij {\it et al.} [LHCb Collaboration],
JHEP {\bf 1810}, 167 (2018)
[arXiv:1806.04079 [hep-ex]].

\bibitem{TheALICE:2014dwa} 
B.~B.~Abelev {\it et al.} [ALICE Collaboration],
Phys.\ Rev.\ Lett.\  {\bf 113}, no. 23, 232504 (2014)
[arXiv:1406.7819 [nucl-ex]].


\bibitem{Hencken:2005hb} 
K.~Hencken, G.~Baur and D.~Trautmann,
Phys.\ Rev.\ Lett.\  {\bf 96}, 012303 (2006)
[hep-ph/0506014].

\bibitem{Schafer:2007mm} 
W.~Sch\"afer and A.~Szczurek,
Phys.\ Rev.\ D {\bf 76}, 094014 (2007)
[arXiv:0705.2887 [hep-ph]].




\bibitem{Bertulani:2005ru}
C.~A.~Bertulani, S.~R.~Klein and J.~Nystrand,
Ann.\ Rev.\ Nucl.\ Part.\ Sci.\  {\bf 55} (2005) 271
[nucl-ex/0502005].

\bibitem{Baur:2001jj}
G.~Baur, K.~Hencken, D.~Trautmann, S.~Sadovsky and Y.~Kharlov,
Phys.\ Rept.\  {\bf 364} (2002) 359
[hep-ph/0112211].

\bibitem{Nikolaev:1993th} 
N.~N.~Nikolaev and B.~G.~Zakharov,
Z.\ Phys.\ C {\bf 64}, 631 (1994)
[hep-ph/9306230].

\bibitem{Chen:2016dlk} 
G.~Chen, Y.~Li, P.~Maris, K.~Tuchin and J.~P.~Vary,
Phys.\ Lett.\ B {\bf 769}, 477 (2017)
[arXiv:1610.04945 [nucl-th]].

\bibitem{Nikolaev:2006mh} 
N.~N.~Nikolaev, W.~Sch\"afer, B.~G.~Zakharov and V.~R.~Zoller,
JETP Lett.\  {\bf 84}, 537 (2007)
[hep-ph/0610319].





\end{thebibliography}
\end{document}